\documentclass[
twocolumn,
pre,
aps,
superscriptaddress,
showpacs,
preprintnumbers,
amsmath,
amssymb]{revtex4-1}
\usepackage{bm}

\usepackage[dvipdfmx]{graphicx}
\graphicspath{{./YAS_picture/}}

\def\erf{{\rm erf}}

\def\picturewidth{8.6cm}%linewidth=246pt=86.7833mm

\begin{document}

\title{Microscopic instability in recurrent neural networks}

\author{Yuzuru Yamanaka}
\email{y.yamanaka@scphys.kyoto-u.ac.jp}
\affiliation{Department of Physics, Kyoto University, Kyoto 606-8502, Japan}
\author{Shun-ichi Amari}
\email{amari@brain.riken.jp}
\affiliation{RIKEN Brain Science Institute, Hirosawa 2-1, Wako-shi Saitama 351-0198, Japan}
\author{Shigeru Shinomoto}
\email{shinomoto@scphys.kyoto-u.ac.jp}
\affiliation{Department of Physics, Kyoto University, Kyoto 606-8502, Japan}

\date{\today}

\begin{abstract}
In a manner similar to the molecular chaos that underlies the stable thermodynamics of gases, neuronal system may exhibit microscopic instability in individual neuronal dynamics while a macroscopic order of the entire population possibly remains stable. In this study, we analyze the microscopic stability of a network of neurons whose macroscopic activity obeys stable dynamics, expressing either monostable, bistable, or periodic state. We reveal that the network exhibits a variety of dynamical states for microscopic instability residing in given stable macroscopic dynamics. The presence of a variety of dynamical states in such a simple random network implies more abundant microscopic fluctuations in real neural networks, which consist of more complex and hierarchically structured interactions.
\end{abstract}

\maketitle

\section{Introduction}

While an animal is repeating a fixed action in response to a given stimulus, individual neurons in the brain do not necessarily reproduce identical activity~\cite{Tolhurst83,Softky93,Churchland11}. The contrast between the reliable animal behavior and the erratic activity of single neurons may be compared with thermodynamics of gases, in which macroscopic states obey thermodynamic laws with small degrees of freedom, while individual molecules obey chaotic dynamics involving large degrees of freedom. In the thermodynamics, the difference in stability is resolved in such a way that macroscopic thermodynamic laws are deduced through the Boltzmann equation describing the microscopic chaotic motion of simplistic model molecules~\cite{Hashitsume91}. It has been one of key objectives of statistical physics to relate microscopic dymanics with macroscopic dynamics in diverse phenomena, including populations of active elements~\cite{Boccaletti06,Strogatz03,Ito01,Kinzel09,Cohen12,Winkler12} and human activity~\cite{Barabasi05,Sornette04,Crane08,Castellano09,Coolen05,Masuda04}.

For networks of randomly connected neurons, it has been shown that a macroscopic order parameter represented by the average neuronal activity obeys deterministic dynamics; in particular, networks of McCulloch--Pitts binary neurons~\cite{bib:McCullochPitts} exhibit three distinct types of macroscopic dynamics, expressing either monostable, bistable, or periodic state~\cite{Amari71}. By contrast, in the same system, it was revealed that a microscopic state specified by a set of individual neuronal states may become unstable against microscopic perturbations, such as flipping a single neuron state~\cite{Shinomoto86}. However, in that study, the microscopic instability was verified solely by numerically simulating small systems, and accordingly, the mechanism of the instability was not examined thoroughly.

Here, we study the microscopic instability of neural networks in detail using analytical as well as numerical analysis, and reveal various types of microscopic dynamics residing in stable macroscopic dynamics.

\section{Random neural network}
\label{sec:RNN}

\begin{figure}[ht]
\begin{center}
\includegraphics[width=\picturewidth]{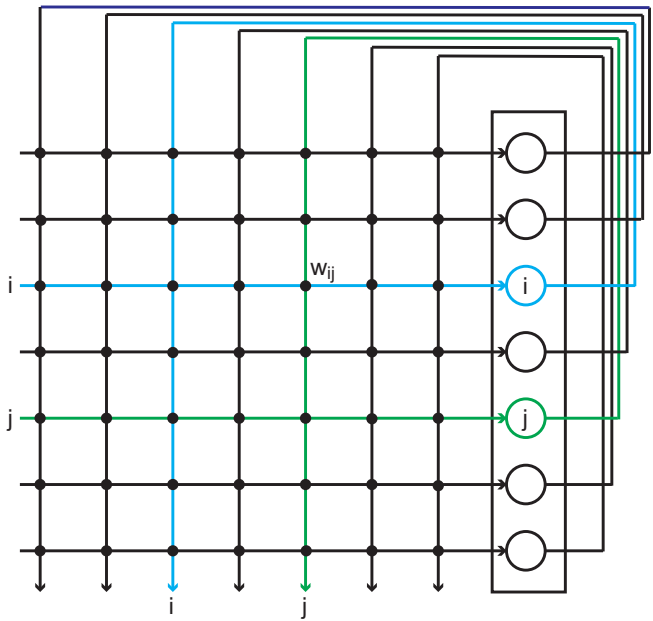}
\caption{A recurrent network of neurons. $w_{ij}$ represents a synaptic connection from the $j$th neuron to the $i$th neuron. Neurons update their states according to Eq.(\ref{evolution}).}
\label{figrandomnet}
\end{center}\end{figure}

We consider a network of McCulloch--Pitts binary neurons interacting via random synaptic connections. Here, we adopt a symmetric expression with active and inactive states respectively represented as
\begin{eqnarray}
s_i(t)= \left \{
\begin{array}{l}
+1,\\
-1,
\end{array}
\right.
\end{eqnarray}
where $i (=1, 2, \cdots, N)$ is the label of a neuron and $t$ is the discretized time given by an integer. In every time step, all the states of an entire neuronal population are updated synchronously, such that each neuron is either activated or deactivated depending on whether the summed input exceeds the threshold or not: 
\begin{eqnarray}
s_i(t+1)= {\rm sgn}{\left( \sum_{i=1}^N w_{ij} s_j(t) + h_i \right)}, 
\label{evolution}
\end{eqnarray}
where ${\rm sgn}(x)$ is the sign function, $N$ is the total number of neurons, $w_{ij}$ represents the synaptic connection from the $j$th neuron to the $i$th neuron, and $-h_i$ is the threshold for the $i$th neuron (Fig.~\ref{figrandomnet}). 

The state evolution rule Eq.(\ref{evolution}) is similar to the zero-temperature relaxation dynamics for a spin glass given by the exchange interactions $w_{ij}$ and external fields $h_i$. Characteristics that distinguish our neural network model from spin systems are as follows: (i) updates are synchronous, (ii) connections $w_{ij}$ are generally asymmetric, and (iii) self-connections $w_{ii}$ can be present. We denote synaptic connections as $w_{ij}$ to distinguish from the exchange interactions of spin systems, usually denoted as $J_{ij}$. 

Note that the neuronal states can alternatively be expressed as $1$ and $0$, a straightforward representation of the active and inactive states, respectively. This can be done by transforming as $u_i(t)=(s_i(t)+1)/2$. Accordingly, the state evolution Eq.(\ref{evolution}) may be rewritten as $u_i(t+1)= \theta{\left( \sum_{i=1}^N w_{ij} u_j(t) + T_i \right)}$, where $\theta(x)$ is the Heaviside step function. In this case, the threshold $T_i$ is given by $(- h_i + \sum_{i=1}^N w_{ij})/2$. 

\section{Macroscopic dynamics}

One of the authors has shown that a macroscopic activity of the neural network obeys deterministic dynamics in the limit of a large number of neurons, $N \to \infty$~\cite{Amari71}. Here, we derive the evolution equation of a network whose synaptic connections $w_{ij}$ are drawn independently from an identical Gaussian distribution with a mean of $\bar w/N$ and a variance of $1/N$, whereas the threshold is chosen as a constant $-h_i = -h$.

\subsection{Evolution equation of macroscopic activity}

Consider the situation in which a set of states $\{ s_i (t) \}$ is selected randomly under a given mean activity,
\begin{eqnarray}
m(t) \equiv \frac{1}{N}\sum_{i=1}^N s_i(t).
\end{eqnarray}
If the neuronal states are statistically independent of the synaptic connections $\{w_{ij}\}$, inputs to individual neurons, given as
\begin{eqnarray}
v_i \equiv \sum_{j=1}^N w_{ij}s_j(t) + h, 
\label{ithinput}
\end{eqnarray}
are expected to distribute normally with a mean of $\bar w m(t) + h$ and a variance of $1$. Thus, the total number of neurons that will be activated in the next step will be
\begin{eqnarray}
Q=\frac{N}{\sqrt{2 \pi}} \int_0^{\infty} dv \exp{\left( - \frac{(v-\bar w m(t)-h)^2}{2} \right)},
\end{eqnarray}
with possible fluctuations in $O(\sqrt{N})$. In the limit of a large number of neurons, $N \to \infty$, the activity level in the next step is determined as $m(t+1)=2Q/N-1$ in terms of the current activity level $m(t)$, thus forming the evolution equation,
\begin{eqnarray}
m(t+1) = {\rm erf}{\left(\frac{\bar w m(t)+h}{\sqrt{2}}\right)},
\label{macrodynamics}
\end{eqnarray}
where ${\rm erf}(x)$ is the error function defined by
\begin{eqnarray}
{\rm erf}(x) \equiv \frac{2}{\sqrt{\pi}} \int_0^x dv e^{-v^2}.
\end{eqnarray}

\begin{figure}[ht]
\begin{center}
\includegraphics[width=\picturewidth]{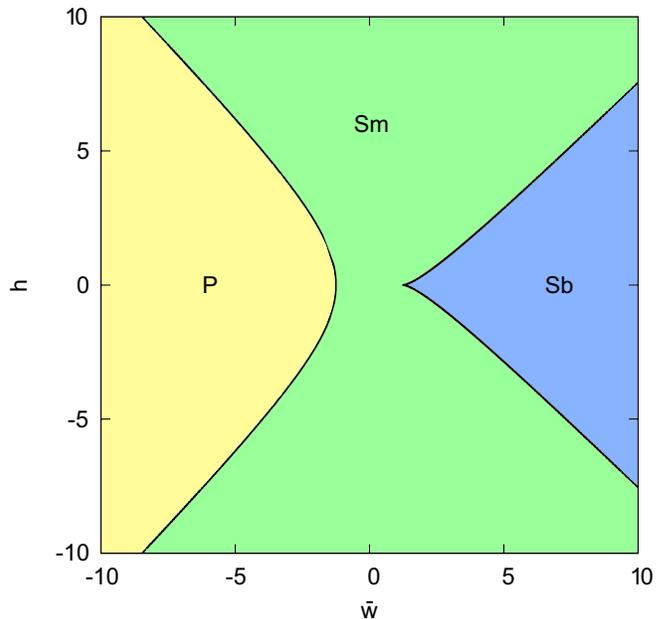}
\caption{Phase diagram of the macrodynamics in a plane of $\bar{w}$ and $h$. P: periodic state, Sm: monostable state, Sb: bistable state.}
\label{figmeanactivity}
\end{center}\end{figure}

Because neuronal states in the next step $\{ s_i (t+1) \}$ are determined by the set of synaptic connections $\{w_{ij}\}$, they are generally not independent of the connections. Nevertheless, the network activity keeps following the evolution of Eq.(\ref{macrodynamics}) if the network is of a reasonably large size~\cite{Amari71}. The evolution equation may show three types of dynamics depending on macroscopic parameters $\bar w$ and $h$: monostable (Sm), bistable (Sb), or periodic (P) states (Fig.~\ref{figmeanactivity}).

\begin{figure}[ht]
\begin{center}
\includegraphics[width=\picturewidth]{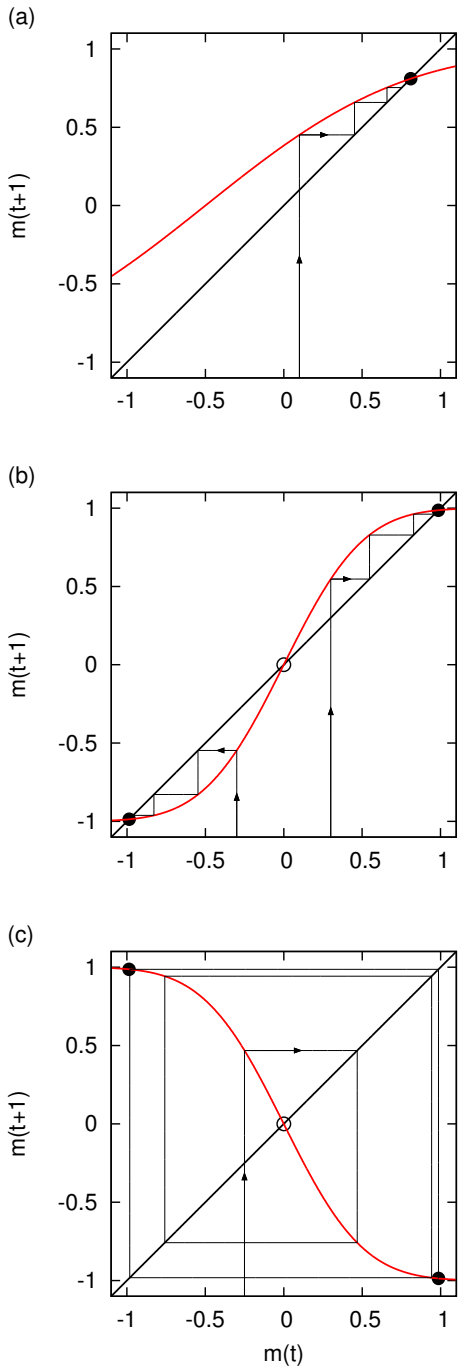}
\caption{Examples of the three types of dynamics of the macroscopic activity: (a) monostable $(\bar w=1,\,h=0.5)$, (b) bistable $(\bar w=2,\,h=0)$, and (c) periodic state $(\bar w=-2,\,h=0)$. }
\label{figmeanactivity_map}
\end{center}\end{figure}

\subsection{Stability of macroscopic dynamics}

The evolution equation Eq.(\ref{macrodynamics}) may have a fixed point $m(t)=m$ that satisfies the self-consistent equation,
\begin{align}
m={\rm erf}{\left(\frac{f}{\sqrt{2}}\right)},
\label{mself}
\end{align}
where $f$ is the average input given as,
\begin{align}
f \equiv \bar{w} m + h.
\label{averageinput}
\end{align}
The macroscopic activity is stable if the absolute slope of the iteration map Eq.(\ref{macrodynamics}) at the point of intersection with the $y=x$ line is smaller than unity:
\begin{align}
\left| \bar{w} \sqrt{\frac{2}{\pi}}e^{-f^2/2} \right| < 1.
\label{P_approx}
\end{align}
The system is called monostable if the evolution equation has only one stable fixed point (Fig.~\ref{figmeanactivity_map}(a)). 
	
By increasing the mean synaptic connection $\bar{w}$ from the monostable regime, the fixed point loses stability when the slope of the iteration map Eq.(\ref{macrodynamics}) at the intersection becomes greater than $1$. The system then becomes bistable, thorough a pitchfork bifurcation, which involves a pair of stable fixed points appearing on both sides of the destabilized fixed point (Fig.~\ref{figmeanactivity_map}(b)). The boundary between the monostable and bistable regimes is obtained by solving $\bar{w} = \sqrt{\pi/2}\exp{(f^2/2)}$ with $m$ satisfying Eq.(\ref{mself}).

Contrariwise, by decreasing $\bar{w}$ from the monostable regime, the single fixed point loses stability when the slope of the iteration map Eq.(\ref{macrodynamics}) at the intersection becomes smaller than $-1$. The system then begins to oscillate through a period-doubling bifurcation (Fig.~\ref{figmeanactivity_map}(c)). In this periodic state, the macroscopic activity $m(t)$ oscillates between the two newly appeared stable fixed points of the iterated map,
\begin{eqnarray}
m=\erf\left(\frac{\bar w\,\erf\left(\frac{\bar wm+h}{\sqrt{2}}\right)+h}{\sqrt{2}}\right).
\end{eqnarray}

\section{Microscopic dynamics}

While the macroscopic order parameter $m(t)$ exhibits stable dynamics following the simple iteration map Eq.(\ref{macrodynamics}), it is possible that a set of neuronal states $\{ s_i (t) \}$ are dynamically changing in time within the given constraint, $m(t) = (1/N)\sum_{i=1}^N s_i(t) \pm O(1/\sqrt{N})$. One of the authors has numerically examined the possibility that the system may be microscopically unstable due to the state flipping of one neuron~\cite{Shinomoto86}. In the present study, we analytically estimate the parameter range of the microscopic instability.

\subsection{Microscopic instability in the macroscopically stable regimes}

We first examine the microscopic stability of the macroscopically stable regime, including the monostable and bistable states. While microscopic states evolve with the individual neuronal dynamics of Eq.(\ref{evolution}), we consider flipping a single neuron state and examine whether the flipping spreads over the network or not. By flipping the state of the $p$th neuron, input to the $i$th neuron is altered from $v_i = \sum_{j=1}^{N} w_{ij}s_j+h $ to  
\begin{align}
v_i^p = \sum_{j=1}^{N}(-2\delta_{jp}+1)w_{ij}s_j+h,
\end{align}
where $\delta_{jp}$ is the Kronecker delta. The state of the $i$th neuron will be altered in the next step if the sign of the input is reversed:
\begin{align}
v_i v_i^p = \left(\sum_{j\neq p}^N w_{ij}s_j +h\right)^2-w_{ip}^2<0.
\label{flipcond}
\end{align}
Under the assumption that $\{w_{ij}\}$ and $\{s_j\}$ are independent, the probability $P$ at which the above mentioned inequality holds is obtained analytically. Because the first and the second terms in the RHS of Eq.(\ref{flipcond}) are distributed normally, the probability $P$ is given as
\begin{align}
P&=
\frac{N}{2\pi\sqrt{N-1}}
\int_{|x|>|y|} dxdy\:
\exp\left( -\frac{N\left(x-\frac{\bar w}{N}\right)^2}{2}\right)\times\nonumber\\
&\hspace{2.7cm}\exp\left(	
-\frac{N\left(y-f+\frac{\bar w s_p}{N}\right)^2}
{2\left(N-1\right)}\right).
\end{align}
In the limit of a large number of neurons, $P$ is approximated as
\begin{align}
P &\approx \frac{1}{2}\sqrt{\frac{N}{2\pi}}\int_{-\infty}^{\infty}dx\:
\exp\left( -\frac{Nx^2}{2}\right) \times \nonumber
\\
&\hspace{0.5cm}\left(\erf\left(\frac{|x|-f}{\sqrt{2}}\right)
-\erf\left( \frac{-|x|-f}{\sqrt{2}}\right)\right)\label{int_P}
\\
&\approx
\frac{2}{\pi\sqrt{N}}
\exp\left( 
-\frac{f^2}{2}
\right).\label{probability}
\end{align}
The system is microscopically unstable if the flipping spreads from a single neuron to more than one neurons, i.e., if $NP>1$. This implies that the microscopic state remains unstable under a given stable macroscopic order. This instability condition is summarized in terms of average input $f \equiv \bar{w} m + h$ as
\begin{align}
|f| \leq I_c \equiv \sqrt{2\log\left(\frac{2\sqrt{N}}{\pi}\right)}.
\label{microscopicinstability}
\end{align}

\begin{figure}[ht]
\begin{center}
\includegraphics[width=\picturewidth]{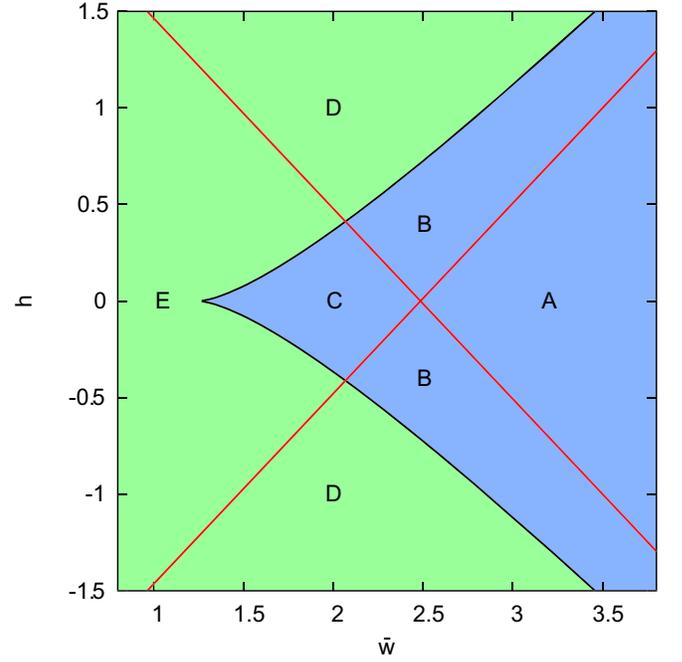}
\caption{Microscopic stability of the macroscopically stable states ($N=1000$). A: Both macroscopically stable states are microscopically stable. B and B': One of two macroscopically stable states is microscopically unstable. C: Both macroscopically stable states are microscopically unstable. D and D': The macroscopically monostable state is microscopically stable. E: The macroscopically monostable state is microscopically unstable. }
\label{figStableRegion}
\end{center}\end{figure}

With this condition, the macroscopically monostable regime can further be classified into two regimes on the basis of whether the system is microscopically unstable or stable. In the bistable regime, in which the system may perform an alternative mean activity, the microscopic instability of the system depends on the macroscopic state. Thus, the bistable parameter regime can be further classified into four regimes on the basis of whether individual macroscopic states are microscopically unstable or stable. The categorized areas are depicted in Fig.~\ref{figStableRegion}.

\subsection{Microscopic instability in the macroscopically periodic regime}

Next, we examine the microscopic instability of the system whose macroscopic activity is oscillating with period two. In the first half of the period two, $m$ is mapped to $ m^\prime=\erf\left(f/\sqrt{2}\right)$ through mean inputs $f\equiv \bar wm+h$. In the second half, $m^\prime$ returns to $m$ through mean inputs $g\equiv \bar w m^\prime +h$ as $ m=\erf\left(g/\sqrt{2}\right)$. Accordingly, $g$ and $f$ are mutually bounded as 
\begin{align}
g=\bar w\,\erf(f/\sqrt{2})+h, 
\label{eqn:g}\\
f=\bar w\,\erf(g/\sqrt{2})+h.
\label{eqn:f}
\end{align} 
It follows from Eqs.(\ref{eqn:g}) and (\ref{eqn:f}) that
\begin{align}
\bar w&=\frac{f-g}{\erf(g/\sqrt{2})-\erf(f/\sqrt{2})}, \label{eqn:wfg}
\\
h&=\frac{f\,\erf(f/\sqrt{2})-g\,\erf(g/\sqrt{2})}{\erf(f/\sqrt{2})-\erf(g/\sqrt{2})}.\label{eqn:hfg}
\end{align}

\begin{figure}[ht]
\begin{center}
\includegraphics[width=8.6cm]{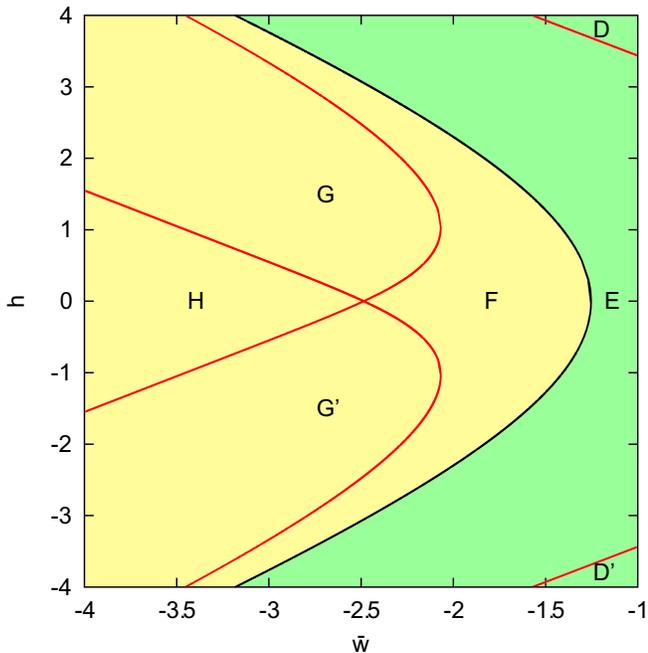}
\caption{Microscopic stability of macroscopically periodic states ($N=1000$). F: Both macroscopic states are microscopically unstable. G and G': One of two macroscopic states is microscopically unstable. H: Both macroscopic states are microscopically stable.}
\label{figPeriodicRegion}
\end{center}
\end{figure}

The system exhibits microscopic instability if both average inputs $|f|$ and $|g|$ are smaller than the critical value $I_c \equiv \sqrt{2\log\left(\frac{2\sqrt{N}}{\pi}\right)}$. The microscopically unstable regime of parameters $(\bar w, h)$ is obtained by searching them under the constraints $|f|<I_c$ and $|g|<I_c$ in Eqs.(\ref{eqn:wfg}) and (\ref{eqn:hfg}). In addition to this perfect instability regime, there are regimes in which either of the two states $m$ and $m^\prime$ is unstable, such that $|f|<I_c<|g|$ or $|g|<I_c<|f|$. The categorized regimes are depicted in Fig.~\ref{figPeriodicRegion}.

\begin{figure}[ht]
\begin{center}
\includegraphics[width=\picturewidth]{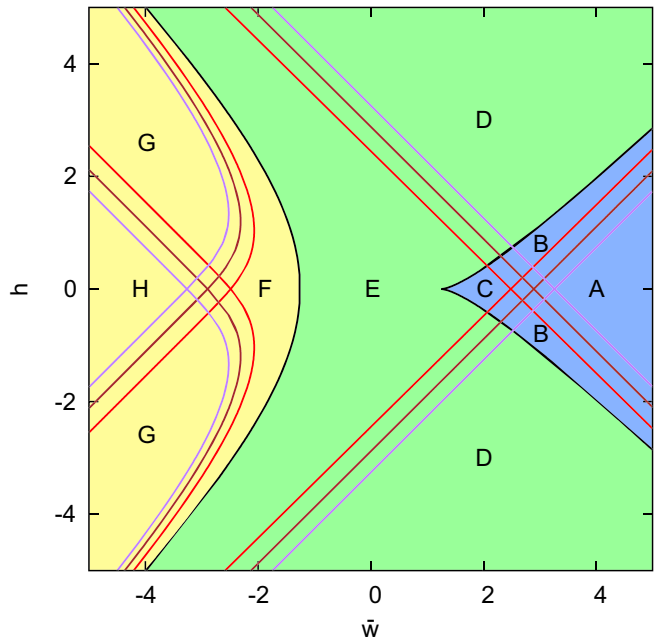}
\caption{Boundaries of microscopic stability. The boundaries for $N$=1000, 10000, and 100000 are depicted in red, brown, and purple, respectively. Regions A to H correspond to the respective regions given in Figs.~\ref{figStableRegion} and \ref{figPeriodicRegion}.}
\label{microscopicboundary}
\end{center}\end{figure}

Notably, the microscopic instability defined by the stability against a one-neuron flip is dependent on the number of neurons $N$, as in Eq.(\ref{microscopicinstability}), and the microscopic instability region may expand without bound. However, the dependence follows the square root of a logarithm, $\sqrt{\log{N}}$, and accordingly, the instability range stays in a small range even in a large network consisting of $O(10^3) - O(10^6)$ neurons (Fig.~\ref{microscopicboundary}).

\section{Numerical Simulation}

A variety of microscopically unstable phases revealed by the current analytical consideration were not observed by the previous simulations of small networks, which ranged from $N=20$ to $200$~\cite{Shinomoto86}. The advancement of computers in recent decades has enabled us to simulate the larger networks. Here, we show the results of simulating a network of size $N=1000$. 

Using a simplified model, we simulated the evolution Eq.(\ref{evolution}) of neurons interacting through synaptic connections distributed normally. Given an initial condition, we iteratively applied the evolution equation for $5000$ steps, expecting that the system would attain macroscopically stable activity. Then the system, starting with the final state, was iterated for one more step. In addition, we flipped a single neuron from the final state and iterated the system for one step. We decide whether the system was microscopically stable or unstable, on the bases of whether the neuronal states of these two systems were entirely identical or not. 

\begin{figure}[ht]
  \begin{center}
\includegraphics[width=\picturewidth]{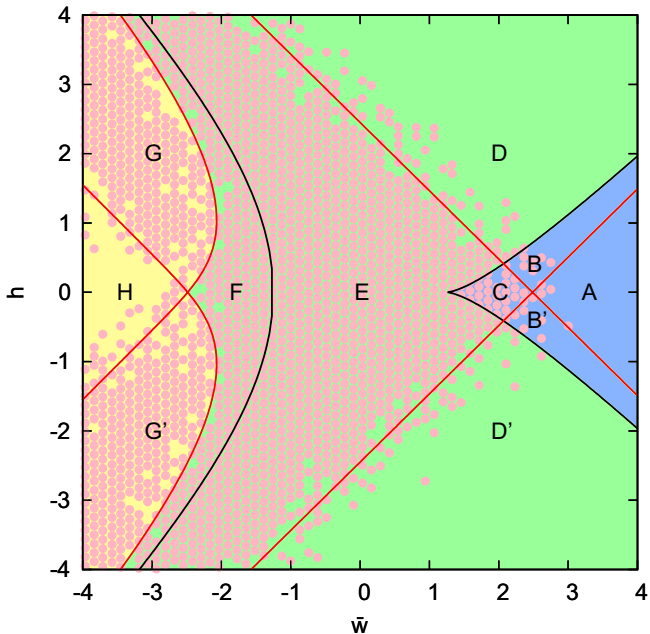}
  \end{center}
  \caption{Microscopic instability obtained by numerical simulation of the networks of $N=1000$. A set of parameters $(\bar{w}, h)$ with which the network exhibit instability is depicted as a dot. Initial states of neurons were chosen at random, $s_i = \pm 1$.}
  \label{fig:M6000}
\end{figure}

\begin{figure}[ht]
  \begin{center}
\includegraphics[width=\picturewidth]{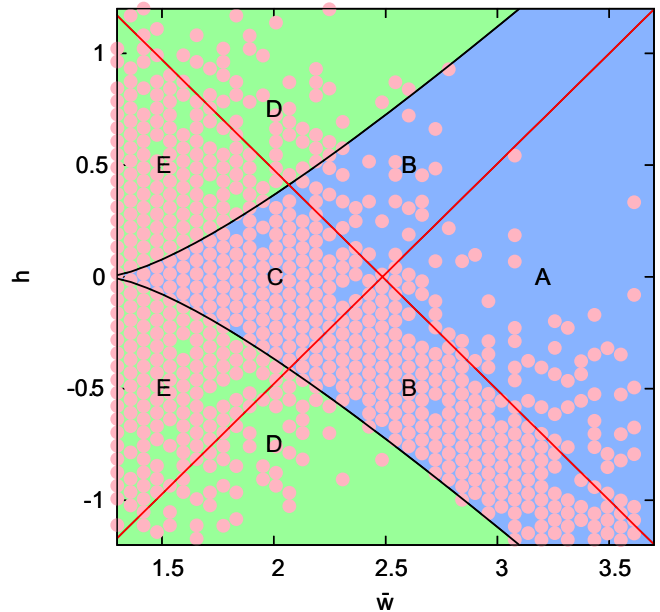}
  \end{center}
  \caption{Microscopic instability obtained from the set of initial states of $\{s_i = +1\}$ ($N=1000$).}
  \label{fig:M1640}
\end{figure}

Figure \ref{fig:M6000} represents the parameters that make the system unstable. In the macroscopically monostable regime, the numerically verified range of microscopic instability is consistent with the theoretical range. In the bistable regime, microscopic stability depends on an alternative macroscopic state. The macroscopic state may be suitably selected by choosing a proper initial condition. For instance, a macroscopic state of higher activation would likely be selected if we choose the initial condition with all the neurons activated, $\{s_i=1\}$. Figure \ref{fig:M1640} depicts the microscopically unstable regime obtained from this all-active initial condition, which is consistent with the analytical result. 

In the periodic regime, microscopic instability may depend on the timing of flipping occurrences. The system remains stable if a neuron is flipped in the stable phase of the oscillation. The microscopic states may deviate if a neuron is flipped in the unstable phase, but these systems may merge in the next step. The numerically obtained unstable regime is similar to the range in which both states $m$ and $m^\prime$ are unstable.

\section{Evolution of the microscopic distance between two states}

Dynamical aspects of microscopic states may also be captured by analyzing the evolution of the distance between two microscopic states. The map of the distance has been obtained for a specific case of $\bar{w}=0$ and $h=0$~\cite{Amari74,Amari13,Derrida86a,Derrida86b,Derrida87}. Here we extend the analysis to general cases of arbitrary $\bar{w}$ and $h$.

Consider the situation that the two states 
\begin{align}
{\bf s}^A&=\{s_1^A,s_2^A,\cdots, s_N^A\},\\
{\bf s}^B&=\{s_1^B,s_2^B,\cdots, s_N^B\},
\end{align}
which possess the identical macroscopic activities $m=m^A=m^B$ expressing the identical macroscopic dynamics, obeying Eq.(\ref{macrodynamics}). Because of the huge amount of combinations of the individual neuronal states, the two states can possibly remain microscopically unidentical. We estimate the evolution of the microscopic distance of two states measured with the normalized Hamming distance,
\begin{align}
d \equiv \frac{1}{2N} \sum_{i=1}^{N} |s_i^A-s_i^B|.
\label{microdistance1}
\end{align}

%\subsection{Distance map in the macroscopically stable regimes}

Here, we consider the case in which the macroscopic activity $m=m^A=m^B$ is stable. We assemble all possible microscopic states ${\bf s}^A (t)$ and ${\bf s}^B (t)$ with macroscopic activity is $m$ and that the mutual distance is $d$, and estimate the distribution of the distance in the next time step,
\begin{align}
d^\prime_i &\equiv \frac{1}{2}|s_i^A-s_i^B|,\\
d^\prime &\equiv \frac{1}{2N} \sum_{i=1}^{N} |s_i^A (t+1)-s_i^B (t+1)|,\\
&=\frac{1}{N}\sum_{i=1}^Nd_i^\prime.
\label{microdistance2}
\end{align}
The mean distance in the next step is obtained as a function of the distance in the current step.
\begin{align}
u_i^\alpha&=\sum_{j=1}^{N}w_{ij}s_j^\alpha+h\hspace{0.5cm}\alpha=A,B\,,\\
u&=\frac{u_i^A+u_i^B}{2\sqrt{(1-d)}},\\
v&=\frac{u_i^A-u_i^B}{2\sqrt{d}},\\
\varphi(d) &\equiv \langle d_i^\prime \rangle  ={\rm Prob}[u_i^Au_i^B<0]\\
&={\rm Prob}\left[|u|<\sqrt{\frac{d}{1-d}}\,|v|\right]. \label{eq32}
\end{align}
Under the assumption that $w_{ij}$ and $s_{i}^\alpha$ are independent and $w_{ij}$ are normally distributed, the RHS of Eq.(\ref{eq32}) is obtained in an integral formula as 
\begin{align}
\varphi(d)
&=\frac{1}{2\pi}\int_{|u|<\sqrt{\frac{d}{1-d}}\,|v|}dudv\,\times\\
&\hspace{2.5cm}\exp\left(-\frac{(u-\frac{f}{\sqrt{1-d}})^2+v^2}{2}\right)\\
%&=\frac{1}{2\pi}\int_{-\infty}^{\infty}dv\;\exp\left(-\frac{v^2}{2}\right)\times\nonumber\\
%&\hspace{0.9cm}\int_{-\sqrt{\frac{d}{1-d}}\,|v|}^{\sqrt{\frac{d}{1-d}}\,|v|}du\;\exp\left(-\frac{(u-\frac{f}{\sqrt{1-d}})^2}{2}\right)\\
&=\frac{1}{2\sqrt{2\pi}}
\int_{-\infty}^{\infty}dv\exp\left(-\frac{v^2}{2}\right)\times\\
&\left\{\erf\left(\frac{\sqrt{d}\,|v|-f}{\sqrt{2(1-d)}}\right)-\erf\left(\frac{-\sqrt{d}\,|v|-f}{\sqrt{2(1-d)}}\right)\right\}.
\end{align}
Note that the microscopic distance is bounded as $d\leq d_{max}\equiv 1-|m|$ for $m=m^{A}=m^{B}$. 

\begin{figure}[ht]
  \centering
  \includegraphics[width=\picturewidth]{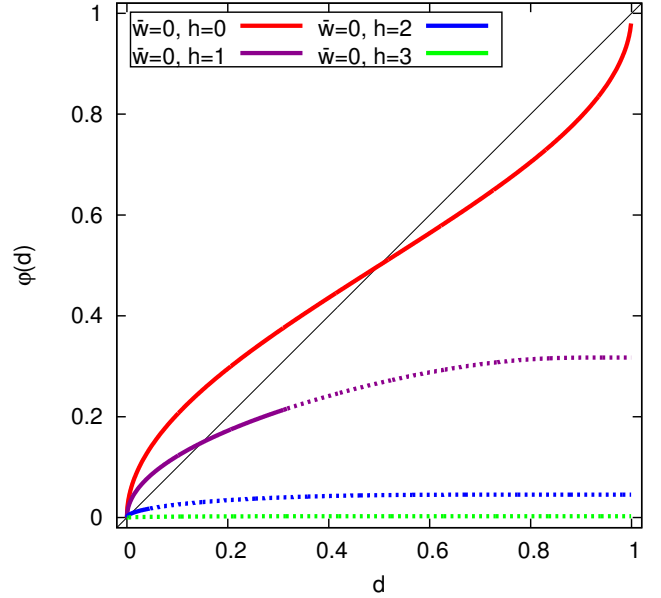}
  \caption{Distance maps, $\varphi(d)$ for various $h$ ($\bar{w}=0$).}
  \label{fig:map.eps}
\end{figure}

Figure \ref{fig:map.eps} represents the evolution map of the distance $\varphi(d)$ for several values of $h$ with $\bar{w}$ kept ar a value of $0$. The map of the case $\bar{w}=h=0$ is $\varphi(d)=\frac{2}{\pi}\sin^{-1}\sqrt{d} $, as has been obtained in Ref.\cite{Amari74,Amari13}.

\begin{figure}[ht]
\centering
   \includegraphics[width=\picturewidth]{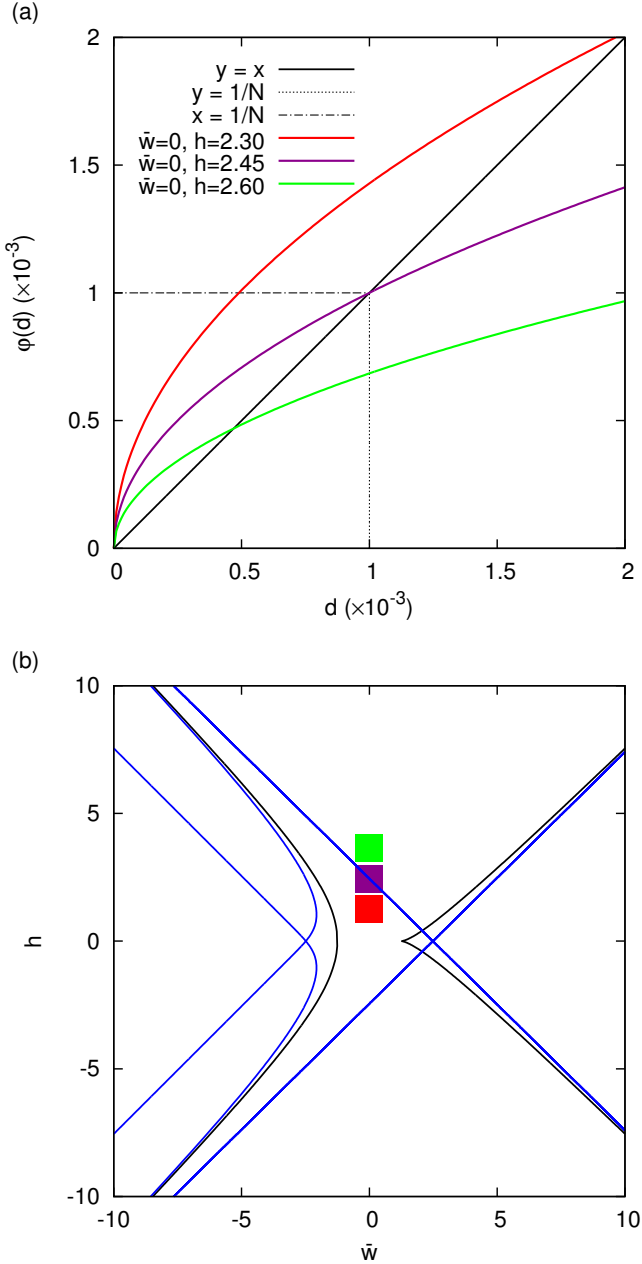}
  \caption{Change in the distance map $\varphi(d)$ across the microscopic stability line ($N=1000$). (a) Three kinds of distance maps. (b) Model parameters for the three distance maps.}
  \label{fig:parameter2.eps}
\end{figure}

When $\sqrt{d}<<f$,
\begin{align}
\varphi (d)&\approx\frac{1}{\pi}\int_{-\infty}^{\infty}dv\,\sqrt{d}\,|v|\exp\left(-\frac{v^2+f^2}{2}\right)\\
&=\frac{2\sqrt{d}}{\pi}\exp\left(-\frac{f^2}{2}\right)
\end{align}
The microscopic distance of $d=1/N$ corresponds to flipping a single neuron. In this case, the average distance in the next step $\varphi(1/N)$ represents the probability of any other single neuron flipping due to the first single neuron flipping. Thus, the condition for microscopic instability discussed in the last section, $NP>1$, is identical to the condition of  
\begin{align}
\varphi\left(\frac{1}{N}\right)\geq\frac{1}{N}.
\label{mapinstability}
\end{align}
Figure \ref{fig:parameter2.eps} represents the manner in which the mapping $\varphi(d)$ varies as the parameters cross the microscopic instability line. It was found from the distance mapping $\varphi(d)$ that the instability of this system is not the linear instability in which the gradient of the mapping exceeds unity, but is simply caused by the inequality Eq.(\ref{mapinstability}).

%\subsection{Distance map in the macroscopically periodic regime}

\section{The period of the microscopic state attractor}

\begin{figure}[ht]
  \begin{center}
   \includegraphics[width=\picturewidth]{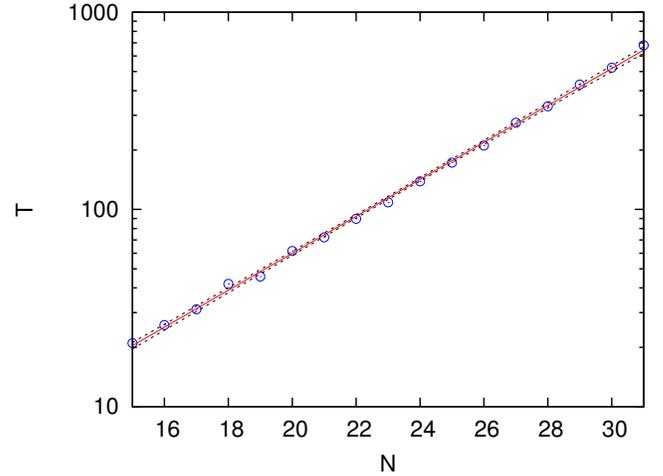}
  \end{center}
  \caption{Period of microscopic states and the number of neurons.}
  \label{fig:Period}
\end{figure}

Because the total number of microscopic states is of finite $2^N$, and the dynamics of individual neurons, described in Eq.(\ref{evolution}), are deterministic, the system eventually enters a cyclic orbit. It has been numerically determined that the period of the attractor cycle of the random neural network of $(\bar{w},h)=(0,0)$ increases exponentially with $N$, on average~\cite{Shinomoto86}. By simulating systems of a size larger than that of the previous study, we confirmed that the logarithm of the periods fits to a linear function of $N$ fairly well (Fig.~\ref{fig:Period}).
\begin{align}
\langle \log{T} \rangle \approx \gamma N + c,
\end{align}
where $\log$ is the natural logarithm. The linear regression analysis applied to the simulation data of $N=15$ to $31$ gives the coefficient $\gamma \approx 0.216 \pm 0.002$. Note that this period is significantly shorter than the average period of the random Boolean map, also called the Kauffman map, which is obtained analytically as $T \sim \sqrt{2^N}$ and the Boolean map's exponent is $\gamma_B = (\log 2 )/2 \approx  0.347 $~\cite{bib:kauffman}. Thus, the period of the microdynamics of the random neural network of $(\bar{w},h)=(0,0)$ typically grows exponentially with $N$, but is shorter than that of the random Boolean map. 

\begin{figure}[ht]
\centering
   \includegraphics[width=\picturewidth]{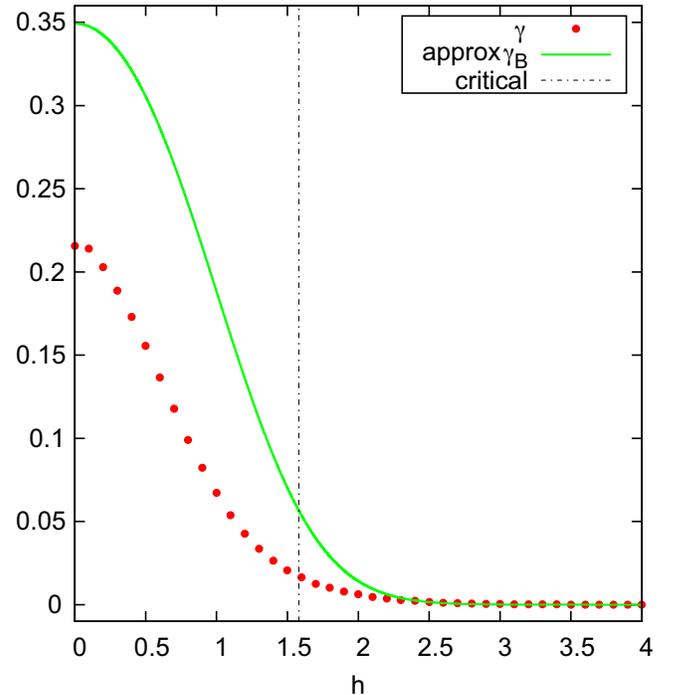}
  \caption{Dependence of the exponent $\gamma$ on the model parameter $h$ ($\bar{w}=0$, $N=30$). $\gamma_B$ is the exponent estimated by the Boolean random map, given a typical number of microscopic states.}
  \label{fig:gamma}
\end{figure}

We now examine how the exponent $\gamma$ changes with the model parameters. Figure \ref{fig:gamma} represents the manner in which the exponent $\gamma$ changes with $h$, while $\bar{w}$ is kept at $0$. In the microscopically stable regime, $|h|>I_c$, where $\varphi (1/N) <1/N$ or the fixed point of the distance map satisfying $\varphi\left( d^* \right) = d^*$ is less than $1/N$, the period of the microscopic attractor $T$ is expected to be unity, implying the average $\log T$ is close to zero. 

Contrariwise, in the microscopically unstable regime, $|h|<I_c$, where the fixed point in the distance map $d^*$ is greater than $1/N$, the microscopic state is expected to meander in state space, and the period of the microscopic attractor $T$ is exponentially large. The number of microscopic states given in the range of the distance $d^*$ is roughly estimated as
\begin{align}
W=\sum_{k=0}^{\lfloor Nd^* \rfloor}{}
\left(
    \begin{array}{c}
      N \\
      k
    \end{array}
  \right),
\end{align}
where $\lfloor x \rfloor$ represents the floor of $x$. When considering the random Boolean map among $W$ states, the typical length of the attractor period is $\sqrt{W}$, and accordingly the exponent $\gamma_B$ is given as
\begin{align}
\gamma_B=\frac{1}{2N}\log{W}.
\end{align}
Figure \ref{fig:gamma} compares the $\gamma_B$ and the real exponent $\gamma$ estimated by numerical simulation. Though the exponent $\gamma_B$ on the basis of the random Boolean map overestimates the actual exponent $\gamma$, the dependence on the model parameter $h$ is qualitatively reproduced.

\section{Discussion}

We have analytically and numerically examined the microscopic dynamics of randomly connected neural networks, and revealed a variety of microscopic dynamics. A network that exhibits stable dynamics in its macroscopic activity may show instability in its microscopic state, as is suggested by the real neural irregular activity in a fixed behavioral context. The analysis of a simplistic system could provide a possible link to the real system. In other words, the real neural network expressing the nonreproducible activity of individual neurons in a fixed behavioral response may represent the microscopic instability in the macroscopic stable dynamics. 

It should be noted that a neural network expresses microscopic instability in the entire parameter region in the limit of a large number of neurons, which corresponds to the thermodynamic limit of gases. Thus the coexistence of microscopic instability with macroscopic stability is expected to play an important role in the information processing in the real neuronal circuitry consisting of a huge number of neurons.

\section*{ACKNOWLEDGMENTS}

We thank Hiromichi Suetani for his stimulating discussions.
This study was supported in part by Grants-in-Aid for Scientific Research to SS from the MEXT Japan (25115718, 26280007) and by JST, CREST.

\end{document}